\documentclass[10pt,conference]{IEEEtran}
\usepackage[letterpaper, left=0.625in, right=0.625in, bottom=1.05in, top=0.75in]{geometry}
\usepackage{epsfig,setspace,amsmath,epsf,amssymb,bm,theorem,cite,graphicx, epstopdf,algorithm,algpseudocode,float,color,mathtools,authblk,tikz,physics}

\usetikzlibrary{positioning}
\usetikzlibrary{decorations.text}
\usetikzlibrary{decorations.pathmorphing}
\usepackage[short,c2]{optidef}

\usepackage{empheq}
\usepackage{cases}

\usepackage{booktabs}
\usepackage{subfigure}
\usepackage{bbm}

\newtheorem{definition}{Definition}

\newcommand{\Exp}{\mathrm{Exp}}
\newcommand{\AoS}{\text{AoS}}

\IEEEoverridecommandlockouts
\allowdisplaybreaks
\begin{document}
\title{A New Metric: Age of Staleness}
\title{Incorrect but Less Stale}
\title{Measuring the Freshness of the Stale Information: Age of Staleness}
\title{Measuring the Freshness of the Outdated Estimation: Age of Staleness}
\title{Maintaining the Most Recent Estimation: Analysis and Optimization of Age of Staleness}
\title{Maintaining the Most Recent Estimation in Digital Twins: Age of Staleness}
\title{When was the Current Estimation Correct: Age of Staleness}
\title{How Far Back in Time the Current Estimation Reflects the Observed Process: Age of Staleness}
\title{A New Synchronization Metric for Digital Twins: Age of Staleness}
\title{How Far Back in Time a Digital Twin Reflects the State of the Physical Object: Age of Staleness}

\author{Ismail Cosandal \qquad Sennur Ulukus\\
        \normalsize Department of Electrical and Computer Engineering\\
        \normalsize University of Maryland, College Park, MD 20742\\
        \normalsize  \emph{ismailc@umd.edu} \qquad \emph{ulukus@umd.edu}}

\maketitle

\begin{abstract}
    The groundbreaking metric age of information (AoI) has been introduced to measure information freshness in communication networks. As transformational as it is, AoI metric falls short in some applications, such as remote monitoring, since it is a semantic-agnostic metric which does not consider the dynamics of the random process. There is a need to quantify the performance of a remote estimator via a metric that combines freshness and semantic aspects. To this end, in this paper, we introduce a novel metric coined \emph{age of staleness (AoS)} that measures when the last time that the current estimation was correct. First, we analyze a simple scenario where an $n$-ary symmetric Markov source is observed by a monitor via a constant sampling rate, obtain a closed-form expression for the AoS, and show that it is a monotonically decreasing function of the sampling rate. Next, we consider multiple distinct Markov sources, and formulate an optimization problem, where the remote monitor allocates the total sampling rate to tracking the sources. Although the optimization problem is non-convex, its structure is suitable for obtaining a near-optimal solution using the \emph{polyblock algorithm}, which leverages the monotonicity of the objective function. While the new AoS metric could be applicable in many scenarios, we believe it is particularly well-suited for a digital twin network (DTN) where multiple physical objects (POs) are monitored with a total sampling rate constraint to maintain a digital representation of them, namely, their \emph{digital twin} (DT). 
\end{abstract}

\section{Introduction}
In remote estimation problems, maintaining fresh information is key to timely and accurate estimation. The age of information (AoI) metric was introduced to quantify the freshness of information \cite{Yates__HowOftenShouldone} and has been used in a variety of problems \cite{yates-survey,kaswan2025age,kahraman2023age}. However, AoI is semantic-agnostic since it only measures the freshness of information, and does so via the time elapsed since the generation time of the latest received update, regardless of the meaning or the purpose of the update itself \cite{maatouk2020}. In  \cite{shisher2024monotonicity, shisher2024timely, ornee2023context}, it is shown that, the estimation performance does not improve monotonically with AoI, for all source types or applications. In \cite{maatouk2020}, it is argued that, an update may still be fresh after a long time, if the source process has not changed after the transmission of the update. Thus,  \cite{maatouk2020} proposes a semantic-aware freshness metric, coined age of incorrect information (AoII), that measures how long an incorrect estimation has been in the system. In various scenarios \cite{maatouk2022age, cosandal_etal_TRIT24, cosandal2025drph}, it is proven that minimizing AoI does not minimize AoII, instead a policy considering the dynamics of the source process is required to reach an accurate and timely information. Even though AoII may be a better metric than AoI in remote estimation problems, AoII only penalizes the incorrect estimation by comparing the current estimate at the remote monitor with the source's current value, and it keeps increasing so long as these two values are different, but, it does not tell anything about how lagging the current estimation is, i.e., when was the last time the current estimation was correct, in other words, \emph{how stale is the current incorrect information}.

\begin{figure}
    \centering
    \vspace{0.1cm}
    \includegraphics[width=0.65\linewidth]{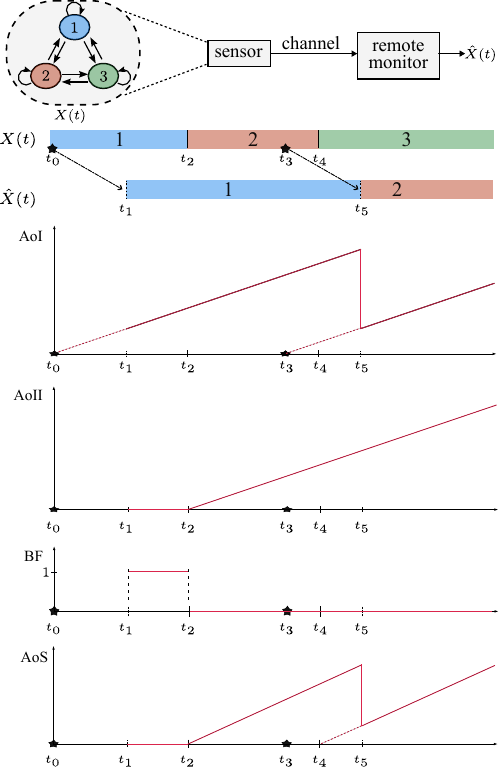}
    \caption{Comparison of AoI, AoII, BF and the proposed AoS metrics for the same $X(t)$ and $\hat{X}(t)$ pair.}
    \label{fig:metrics}
    \vspace*{-0.3cm}
\end{figure}

In remote estimation problems, changes in the source process are reflected with delays, hence, the estimate is always lagging behind the original process. However, as mentioned, the existing metrics do not directly reflect the amount of this lag. We illustrate this with an example process $X(t)$ and its estimation $\hat{X}(t)$ in Fig.~\ref{fig:metrics}. The process is initiated at $t_0$ with state $1$, and the state of the process changes at $t_2$ and $t_4$. A sensor samples the process at $t_0$ and $t_3$, and these samples arrive at the monitor at $t_1$ and $t_5$. AoI is defined as
\begin{align}
    \text{AoI}(t)=t-u(t), \label{eq:AoI}
\end{align}
where $u(t)$ is the generation time of the latest received sample at $t$. Since AoI is a semantic-agnostic freshness metric, it does not say anything about the staleness of the estimation. Binary freshness (BF), on the other hand, is a basic semantic-aware metric that only indicates the accuracy of the estimation by taking the value of $1$ when the current estimate is correct, and $0$ otherwise \cite{bastopcu2021,cosandal2023timely}. Meanwhile, AoII is both a semantic and a freshness-aware metric, formulated as
\begin{align}
    \text{AoII}(t)=t-\max(t'|X(t')=\hat{X}(t')), \label{eq:AoII}
\end{align}
that measures how long the estimation process has been incorrect. Notice that, in Fig.~\ref{fig:metrics}, after $t_5$, the estimate changes from $1$ to $2$. Even though both estimations are incorrect, the estimate $2$ represents relatively more fresh (or less stale) information about the original process. To this end, we propose a novel metric, coined \emph{age of staleness} (AoS), that measures the staleness level of the current estimate by considering when was the last time that the current estimation was correct. We denote this metric by $\text{AoS}(t)$, which is expressed as
\begin{align}
    \AoS(t)=t-\max(t'|X(t')=\hat{X}(t)). \label{eq:SP}
\end{align}
This metric better-reflects the performance of estimator configurations, where relatively longer delays disallow the remote monitor from synchronizing its estimate perfectly with the original process. Specifically, if the process changes frequently and the delays are relatively large, the monitor may not be able to have the correct information, and the AoII metric may continuously increase going to infinity over time. For such a scenario, we argue that having \emph{less stale incorrect information} indicates better estimation performance, and this is precisely what AoS is able to track while AoII is not. 

In the first part of the paper, we consider a simple remote-estimation scenario in which a source process $X(t)$ is sampled with constant sampling rate $\lambda$, and a remote monitor constructs an estimation process $\hat{X}(t)$ using these samples. Notice that for a binary state process, AoS and AoII are equivalent, and this problems has already been investigated in the literature \cite{kam2020age, maatouk2022age}. Thus, we consider another special and analytically tractable source model commonly used in the literature \cite{maatouk2020, yan2024age, chen2022preempting}, namely, the $n$-ary symmetric Markov process, which can be characterized with two parameters only: holding rate at a state and the number of states. To obtain a closed-form expression for the AoS for this scenario, we utilize the absorbing Markov chain (AMC) formulation \cite{akar2023distribution,  cosandal_etal_TRIT24, cosandal2025drph, scheuvens2021state}. Our analysis indicates that AoS is a non-convex, but monotonically decreasing function of $\lambda$.

In the second part of the paper, we consider the problem of tracking multiple heterogeneous Markov sources. We formulate an optimization problem in which the remote monitor aims to minimize the sum of AoS of the Markov sources, under a total sampling rate constraint. To solve this optimization problem, we utilize the polyblock algorithm, which is known for finding near-optimal solutions in finite iterations for monotonic optimization problems with closed constraint sets \cite{zhang2013monotonic, bjornson2013optimal}. 

The AoS metric proposed AoS in this paper may be useful in any remote estimation problem, but is particularly suited for a digital twin network (DTN) scenario, where multiple physical objects (POs) are monitored with real-time data to obtain a digital twin (DT) representing virtual copies of them \cite{sharma2022digital}. For such purpose, mostly AoI \cite{shu2022age, guo2024age, vaezi2023digital} or AoI-like \cite{loubany2025age, duran2023age} semantic-agnostic freshness metrics and several semantic-aware freshness metrics have been explored \cite{tong2023semantic, quazi2024building}. For such applications, our AoS metric will indicate how far back in time the current DT reflects the states of the POs. 

\section{AoS Analysis}
We obtain the expected value of $\AoS(t)$ for the system model in Fig.~\ref{fig:metrics}. We consider a source process $X(t)$ which is a symmetric continuous-time Markov chain (CTMC) with $n$ states and holding rate $\sigma$. That means that the process remains in each state for an exponentially distributed random time with rate $\sigma$, denoted by $\Exp(\sigma)$, after which the state of the process changes to any other state with uniform probability. Equivalently, the process changes its state to any other state with a constant rate $\frac{\sigma}{n-1}$. The generator of the process can be expressed by an $n \times n$ matrix $\bm{Q}$ those diagonal elements are $-\sigma$, and the non-diagonal elements are $\frac{\sigma}{n-1}$.

We consider that a sensor observes the process $X(t)$, and updates the monitor with a constant update rate. The channel is considered error-free, and the $\ell$th transmission takes $L_\ell\sim\Exp(\lambda)$ duration, where $\lambda$ is the update rate controlled by the monitor. After the transmission is completed, the sensor takes another sample and restarts the transmission. We denote the time when the $\ell$th sample is taken (and when transmission of the $(\ell-1)$th sample is completed) with $T_\ell=\sum_{m=1}^{\ell}L_m$. Until time $T_1$, the monitor has no estimation, and the monitor updates the estimation process $\hat{X}(t)$ after each reception. Mathematically, this rule can be expressed as
\begin{align}
    \hat{X}(t)=X(T_{\ell-1}), \quad T_{\ell}\leq t < T_{\ell+1}.  \label{eq:est}
\end{align}

To analyze the average AoS for this configuration, we divide the infinite horizon into periods, where the $\ell$th period starts with a sampling at $t=T_\ell$, and lasts until the transmission of the sample is completed at $t=T_{\ell+1}$. Furthermore, within this period, we define a continuous-time AMC, $Z(t)$. That process has two transient states in the order of $\{S,A\}$, and a single absorption state corresponds to the end of the period. Transient states $S$ and $A$ correspond synchronization event $X(t)=\hat{X}(t)$. Equivalently $X(t)=X(T_{\ell-1})$ for $T_\ell\geq t>T_{\ell+1}$ by \eqref{eq:est}, and asynchronization event $X(t)\neq\hat{X}(t)$, respectively. 

An AMC is formulated by two parameters $\bm{\beta}_k$ and $\bm{A}_k$, which are called the initial probability vector (IPV) and the transient sub-generator (TSG), respectively \cite{kemeny1960finite}. $Z(t)$ is initiated from one of these transient states with $\bm{\beta}=\begin{bmatrix} \beta_{1} &\beta_{2}\end{bmatrix}$, which can be expressed as
\begin{align}
    \beta_{1}&=\mathbb{P}(Z(0)=S)=\mathbb{P}(X(T_\ell)=X(T_{\ell-1})), \quad \forall\ell, \\ 
    \beta_{2}&=\mathbb{P}(Z(0)=A)=\mathbb{P}(X(T_\ell)\neq X(T_{\ell-1})), \quad \forall\ell, \label{eq:beta_def}
\end{align}
and it is terminated upon reaching the absorbing state. Notice that $\beta_1$ is the probability that the source process $X(t)$ stays in the same state after an exponentially random time $L_\ell\sim\Exp(\lambda)$. Similarly, $\beta_2$ is the same probability for the source to change its state. These can be calculated by obtaining the expected value of Kolmogorov's forward equation \cite{durrett1999essentials} over the random variable $L_\ell$. Therefore, IPV is
\begin{align}
    \bm{\beta}=\begin{bmatrix} \dfrac{\lambda+\rho}{\lambda+\sigma+\rho}, \dfrac{\sigma}{\lambda+\sigma+\rho}\end{bmatrix}. \label{eq:beta_k}
\end{align}

Next, TSG $\bm{A}$ corresponds to the transition rates between the transient states. From the transient state $S$, a transition to the transient state $A$ happens if the process $X(t)$ changes its state to any other state, which happens with rate $\sigma$. Otherwise, it stays in the same state until $Z(t)$ is absorbed with rate $\lambda$, or a state change happens with rate $\sigma$. Therefore, the holding rate for the transient state $S$ is $\lambda+\sigma$, and we denote the $m$th holding duration for the transient state $S$ by $H_{S,m}\sim\Exp(\lambda+\sigma)$. Additionally, from the transient state $A$, transition to $S$ occurs if the process $X(t)$ returns to the estimated state, which occurs with rate $\frac{\sigma}{n-1}$, or it is absorbed with rate $\lambda$. The holding rate for the transient state $A$ is $\lambda+\frac{\sigma}{n-1}$, and we denote the $m$th holding duration by $H_{A,m}\sim\Exp(\lambda+\sigma)$. Therefore, TSG is
\begin{align}
    \bm{A}=\begin{bmatrix}
            -\sigma-\lambda & \sigma \\
            \rho & -\rho-\lambda 
    \end{bmatrix}, \label{eq:A_k}
\end{align}
where $\rho=\frac{\sigma}{n-1}$.

\begin{figure}
    \centering
    \vspace{0.1cm}
    \includegraphics[width=0.99\linewidth]{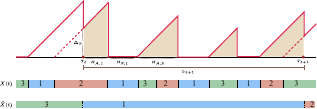}
    \caption{The evolution of the AoS over a single period.}
    \label{fig:period}
    \vspace*{-0.3cm}
\end{figure}

Fig.~\ref{fig:period} illustrates a single period. In the beginning of the period at $t=T_\ell$, the monitor updates its estimate with $X(T_{\ell-1})=1$, thus $\hat{X}(t)=1$ for $T_\ell\leq t < T_{\ell+1}$ by \eqref{eq:est}. For this example, events $X(t)=1$ and $X(t)=2,3$ correspond to transient states $S$ and $A$ of $Z(t)$, respectively. Each period has identical statistical characteristics since the source process $X(t)$ is symmetric. Therefore, analyzing a single period is sufficient to obtain the expected AoS. For a more general process, we can use the \emph{embedded chain representation} in \cite{cosandal_etal_TRIT24}. 

The filled area in the Fig.~\ref{fig:period} represents the accumulated AoS cost for the period. The expected AoS can be expressed by the ratio of the expected value of the accumulated AoS cost to the expected period duration as
    %\vspace*{-0.2cm}
\begin{align}
    \mathbb{E}[\text{AoS}]&=\dfrac{\mathbb{E}\left[ \int_{T_\ell}^{T_{\ell+1}}\text{S}(t)dt\right]}{\mathbb{E}[T_{\ell+1}-T_\ell]} \\
    &=\dfrac{\mathbb{E}\left[\frac{1}{2}\sum_{m=1}^MH_{A,m}^2+\Delta_0 H_{A,1}\right]}{\mathbb{E}[L_{\ell+1}]}\\
    &= \lambda \mathbb{E} \big[ {\textstyle \frac{1}{2}\sum_{m=1}^M} H_{A,m}^2 + \Delta_0 H_{A,1} \big], \label{eq:cost}\end{align}
\noindent{where} $\Delta_0$ is the initial AoS, and $M$ is the number of occurrences of the transient state $A$ over a period.

The $\Delta_0$ is equivalent to the last time the new estimation was correct at $t=T_\ell$, which is expressed as
\begin{align}
    \Delta_0=T_{\ell}-\sup\{t\leq T_{\ell}:X(T_\ell-t)=X(T_{\ell-1}) \} , \label{eq:d0_1}
\end{align}
Let us define a time-reversed process
\begin{align}
    Y(t)=X(T_\ell-t),
\end{align}
where $Y(t)$ has the same generator $\bm{Q}$ by the symmetry. Then, \eqref{eq:d0_1} is equivalent to the first passage time of $Y(t)$ for the state $Y(L_{\ell-1})$, which is expressed as
\begin{align}
    \Delta_0=\sup\{t\leq L_{\ell-1}:Y(t)=Y(L_{\ell-1})\}.
\end{align}
We consider two cases: i) $Y(0)=Y(L_{\ell-1})$, and ii) $Y(0)\neq Y(L_{\ell-1})$. The first case is trivial, which gives us $\Delta_0=0$ with probability $\beta_{1}$ by \eqref{eq:beta_def}. The second case occurs with probability $\beta_{2}$, and it is the minimum of the first passage time of state $X(T_{\ell-1})$, and random time $L_{\ell-1}\sim\Exp(\lambda)$, i.e.,
\begin{align}
    \Delta_0&=\min(\Exp(\rho),\Exp(\lambda)), \ \text{w.p.} \ \beta_{2} \\
    &=\Exp(\rho+\lambda), \ \text{w.p.} \ \beta_{2}.
\end{align}
Finally, we obtain the distribution of $\Delta_0$ as
\begin{align}
    \Delta_0=\begin{cases}
        0, & \text{w.p.} \quad \dfrac{\lambda+\rho}{\lambda+\sigma+\rho}, \\
        \Exp(\lambda+\rho), & \text{w.p.} \quad \dfrac{\sigma}{\lambda+\sigma+\rho},
    \end{cases}
\end{align}
and the expected value of the initial AoS can be calculated as
\begin{align}
    \mathbb{E}[\Delta_0]=\dfrac{\sigma}{\lambda+\sigma+\rho}\dfrac{1}{\lambda+\rho}. \label{eq:Ed0}
\end{align}

A basic property about an AMC is that the expected number of visits to a transient state $j$ starting from a transient state $i$ is given by the $(i,j)$th entry of the so-called fundamental matrix $\bm F=(\bm I-\bm D)^{-1}$ \cite{kemeny1960finite}, where $\bm{D}$ is the embedded Markov chain obtained by $\bm{A}$ as $\bm{D}=\begin{bsmallmatrix}
        0 & \frac{\sigma}{\sigma+\lambda} \\ \frac{\rho}{\rho+\lambda} & 0
    \end{bsmallmatrix}$.
Therefore, the expected visit of transient state $A$, $\mathbb{E}[M]$, is obtained as
\begin{align}
    \mathbb{E}[M]&=\bm{\beta} (\bm I-\bm D)^{-1}\bm{e}_2=\dfrac{\sigma(2\lambda+\rho+\sigma)(\rho+\lambda)}{\lambda(\lambda+\rho+\sigma)^2}, \label{eq:EN}
\end{align}
where $\bm{e}_k$ is a column vector of all $0$s except a $1$ at $k$th position.

The expression in \eqref{eq:cost} can be simplified by using the independence between $H_{A,1}$ and $\Delta_0$, and Wald's identity since i) $\mathbb{E}[H_A]<\infty$, ii) $\mathbb{E}[M]<\infty$, and iii) $M$ is a stopping time independent from the future value of $H_{A,m}$ for $m>M$ \cite{serfozo2009basics}. The simplified equation is
\begin{align}
        \mathbb{E}[\AoS]&=\frac{\lambda}{2}\mathbb{E}\left[ M\right]\mathbb{E}\left[H_{A}^2\right]+\lambda\mathbb{E}\left[\Delta_0\right]\mathbb{E}\left[ H_{A,1}\right], \label{eq:cost2}
\end{align}

Now, we are ready to obtain the expected AoS from \eqref{eq:cost2} by using $\mathbb{E}[H_A^2]=\frac{2}{(\lambda+\rho)^2}$, $\mathbb{E}[M]$ \eqref{eq:EN}, and $\mathbb{E}[\Delta_0]$ \eqref{eq:Ed0} as
\begin{align}
    \mathbb{E}[\AoS]=\dfrac{2\lambda+\rho}{(\lambda+\rho)^2}-\dfrac{2\lambda+\rho+\sigma}
{(\lambda+\rho+\sigma)^2}. \label{eq:ES}
\end{align}
%\newpage
For future comparison, next we derive the average BF. By the definition of BF, the average value is equivalent to the expected duration of the transient state $A$ over the expected duration of a period, and is calculated as
\begin{align}
    \mathbb{E}\left[\text{BF}\right]&=\dfrac{\mathbb{E}\left[ \int_{T_\ell}^{T_{\ell+1}} \mathrm{1}_{X(t)=\hat{X}(t)}dt \right]}{{\mathbb{E}[T_{\ell+1}-T_\ell]}} \\
    &=\dfrac{ \mathbb{E}[N]\mathbb{E}[H_A] }{\mathbb{E}[L_{\ell+1}]} \\
    &=\dfrac{\sigma(2\lambda+\rho+\sigma)}{(\lambda+\rho+\sigma)^2}. \label{eq:BF}
\end{align}

Similarly, next, we derive the average AoI. At time $T_\ell$, the update generated at $T_{\ell-1}$ is received by the monitor, thus AoI reduces to AoI$(T_\ell)=L_\ell$, and it increases to AoI$(T_{\ell+1})=L_\ell+L_{\ell+1}$ by time $T_{\ell+1}$. Thus, the expected value of AoI is calculated as
\begin{align}
    \mathbb{E}[\text{AoI}]&=\dfrac{\mathbb{E}\left[ \int_{T_\ell}^{T_{\ell+1}}\text{AoI}(t)dt\right]}{\mathbb{E}[T_{\ell+1}-T_\ell]} \\
    &=\dfrac{\mathbb{E}[\frac{1}{2}L_{\ell+1}^2+L_\ell L_{\ell+1}]}{\mathbb{E}[L_{\ell+1}]}\\
    &=\dfrac{\frac{1}{\lambda^2}+\frac{1}{\lambda^2}}{\frac{1}{\lambda}}=\frac{2}{\lambda}. \label{eq:AoI}
\end{align}
Thus, the expected AoS, BF and AoI are given in \eqref{eq:ES}, \eqref{eq:BF} and \eqref{eq:AoI}, respectively.

\section{Multiple Sources and Sum AoS Optimization}
\begin{figure}[t]
    \centering
    %\vspace{0.1cm}
    \includegraphics[width=0.75\linewidth]{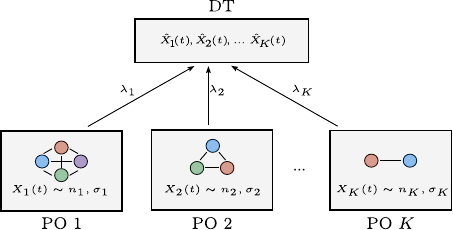}
    \caption{A monitor samples multiple sources. E.g., a DT monitor observes multiple POs with different sampling rates under a total sampling rate budget.}
    \label{fig:DTN}
    \vspace*{-0.3cm}
\end{figure}

We consider the case of multiple sources, where a monitor observes $K$ sources; as an example, this may represent a DT sampling different POs under a total sampling rate constraint as shown in Fig.~\ref{fig:DTN}. We aim to find the optimum sampling rates that satisfy $\sum_{k=1}^K \lambda_k\leq1$ to minimize the sum of the individual AoS values. Here, we use the vector notation with $\bm{a}=[a_1, \dots, a_K]$ representing the collection of parameter $a$ of all sources. Further, $\bm{a}\geq\bm{b}$ indicates the vector $\bm{a}$ is element-wise greater than or equal to the vector $\bm{b}$. We express the objective function of the problem as
\begin{align}
    g(\bm{\lambda};\bm{\sigma},\bm{n})=\sum_{k=1}^K  f_k\left(\lambda_k;\sigma_k, \frac{\sigma_k}{n_k-1}\right),
\end{align}
where $f_k(\lambda_k;\sigma_k,\rho_k)$ is the expected AoS of source-$k$ in \eqref{eq:ES}, 
\begin{align}
    f_k(\lambda_k;\sigma_k,\rho_k)=\dfrac{2\lambda_k+\rho_k}{(\lambda_k+\rho_k)^2}-\dfrac{2\lambda_k+\rho_k+\sigma_k}{(\lambda_k+\rho_k+\sigma_k)^2}. \label{eq:ES1}
\end{align}
Thus, the optimization problem can be expressed as
\begin{mini}
	{ \bm{0}\leq \bm{\lambda} \leq \bm{1}}{g(\bm{\lambda};\bm{\sigma},\bm{n}) }
	{\label{Opt1}}
    {}
	\addConstraint{ \bm{\lambda}\bm{1}^\intercal }{\leq 1,}  
\end{mini}
where $\bm{1}$ and $\bm{0}$ are all-one and all-zero row vectors, respectively. To begin, note that the objective function is a non-convex but monotonically decreasing function for each $\lambda_k$. In addition, the constraint set of the problem, denoted by $\mathcal{G}$, is a normal set, meaning that if vector $\bm{\lambda}$ is feasible, any vector $\hat{\bm{\lambda}}\leq\bm{\lambda}$ is also feasible. 

\subsection{Polyblock Algorithm}
An important property of the monotonic optimization is that the optimum solution lies on the boundary of the constraint set \cite{bjornson2013optimal}, since the monotonicity indicates that 
\begin{align}
    g(\bm{v};\bm{\sigma},\bm{n})\leq g(\bm{u};\bm{\sigma},\bm{n}), \ \text{ if }\bm{v}\geq\bm{u}. \label{eq:mont}
\end{align}
The polyblock algorithm is known to find the near-optimum solution for a monotonic objective function with normal constraint set by iteratively reducing the search space \cite{zhang2013monotonic,bjornson2013optimal}. 

\begin{definition}[Box]
    A box$(\bm{v})$ is defined by a vector $\bm{v}$, which is called \emph{vertex}, that represents the interval $[\bm{0} \ \bm{v}]$. Box$(\bm{v})$ includes any vector $\bm{u}$ satisfying $\bm{0}\leq \bm{u} \leq \bm{v}$.
    \end{definition}

\begin{definition}[Polyblock]
    A polyblock $\mathcal{P}$ is generated by union of boxes. For a vertex set $\mathcal{V}=\{\bm{v}_1,\dots,\bm{v}_n\}$, the polyblock $\mathcal{P}(\mathcal{V})$ is $\mathcal{P}(\mathcal{V})=\text{box}(\bm{v}_1)\cup\text{box}(\bm{v}_1) \dots \cup \text{box}(\bm{v}_n)$.
\end{definition}

In each iteration of the algorithm, we keep a vertex set $\mathcal{V}_i$, and generate a polyblock $\mathcal{P}_i=\mathcal{P}(\mathcal{V}_i)$. We additionally define subsets $\mathcal{V}_i^-$ and $\mathcal{V}_i^+$ that include the vertices from $\mathcal{V}_i$ on the boundary, and outside of the constraint set $\mathcal{G}$, respectively. This way, we generate a new polyblock in each iteration that eventually encloses the constraint set as $\mathcal{P}_1 \supset \mathcal{P}_2 \cdots \mathcal{P}_n \supset \mathcal{G}$.

Polyblock algorithm is initiated with the vertex subsets $\mathcal{V}_1^-=\{\}$ and $\mathcal{V}_1^+=\mathcal{V}_1=\{\bm{v}_u\}$. The vertex $\bm{v}_u$ is called \emph{utopia point}, and for our problem, it is selected as $\bm{v}_u=\bm{1}$. Thus, the first polyblock $\mathcal{P}_1(\mathcal{V}_1)$ is guaranteed to enclose the constraint set $\mathcal{G}$. In each iteration, we first choose a candidate vertex, $\bm{v}'$ from $\mathcal{V}_i^+$ that minimizes the objective function as
\begin{align}
    \bm{v}'=\underset{\bm{v}\in\mathcal{V}_i^+}{\arg\min} \ g(\bm{v};\bm{\sigma},\bm{n}).
\end{align}
Let us denote the optimum solution and its value by $\bm{\lambda}^*$, and $g^*=g(\bm{\lambda^*;\bm{\sigma},\bm{n})}$, respectively. Because the polyblock always includes $\bm{\lambda}^*$, the candidate vertex gives us a lower bound as $lb=g(\bm{\bm{v}';\bm{\sigma},\bm{n})} \leq g^*.$ Similarly, the best solution from $\mathcal{V}_i^-$ gives an upper bound on the problem, which is defined as $ ub=\min_{\bm{v}\in\mathcal{V}_i^-} \ g(\bm{v};\bm{\sigma},\bm{n}) \geq g^*.$

The polyblock algorithm is terminated when $ub-lb$ is smaller than $\epsilon$, and it is known to be reached in finite iterations \cite{zhang2013monotonic}.

Next step, we project $\bm{v}'$ over the constraint set $\mathcal{G}$ by multiplying with a scalar projection coeffient $0\leq\phi\leq1$. We define the projected vertex as $\bm{v}'_\phi$, which is expressed as
\begin{align}
    \bm{v}'_\phi&= \phi^* \bm{v}', \quad  \phi^*=\sup_{0\leq\phi\leq1}\{\phi:\phi\bm{v}'\in\mathcal{G}\}
\end{align}
This operation is equivalent to the normalization of the vector $\bm{v}'$, thus $\bm{v}'_\phi= \bm{v}'/(\bm{1}^{\intercal}\bm{v}')$.

Since there are no feasible points in the interval $[\bm{v}'_\phi,\bm{v}']$, the vertex $\bm{v}'$ is removed from the set $\mathcal{V}_i^+$. Instead, we add the other corners of $[\bm{v}'_\phi,\bm{v}']$ to obtain a smaller polyblock enclosing $\mathcal{G}$. These corners are obtained by replacing the $k$th component of $\bm{v}'$ by $\bm{v}'_{\phi}$ for each $k$, which is equivalent to
\begin{align}
    \bm{v}'_{\phi,k}=\bm{v}'-(\bm{v}'-\bm{v}'_\phi)\bm{e}_k^\intercal \quad k=1,\dots,K.
\end{align}
For the sake of efficiency, we consider two conditions to add these vertices to the vertex set $\mathcal{V}_i^+$. First, as proposed in \cite{zhang2013monotonic}, any vertex $\bm{v}$ whose value is greater than the current $lb$ does not serve the solution, thus should not be added to the vertex set. Additionally, if the candidate vertex is close to any axis, i.e., $v'_k<\delta$ for any $k$ and infinitesimal $\delta$, and $K>2$, one of these newly generated vertices becomes very close to the candidate vertex, and generating this vertex slows the convergence to the solution. Both these conditions can be expressed as $g(\bm{v}'_{\phi,k};\bm{\sigma},\bm{n})<ub$, $||\bm{v}'_{\phi,k}-\bm{v}'||>\delta$, where the operation $||\cdot||$ denotes the Eucledian norm. 

\begin{figure}
    \centering
    \vspace{0.1cm}
    \includegraphics[width=0.65\linewidth]{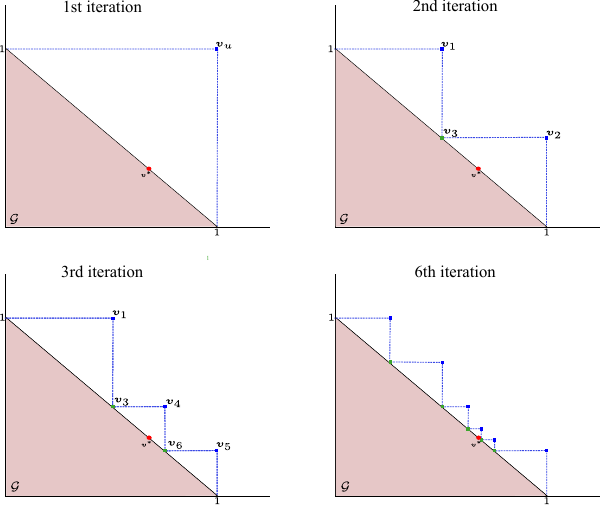}
    \caption{The illustration of how the polyblock algorithm works for $n=2$. Vertices belonging to the subsets $\mathcal{V}^-$ and $\mathcal{V}^+$ are shown with green and blue squares, and the optimum solution $\bm{v}^*$ is marked with a red circle.}
    \label{fig:pa}
    \vspace*{-0.3cm}
\end{figure}

Fig.~\ref{fig:pa} illustrates the first steps of the polyblock algorithm for an example case with $n=2$. The first polyblock is generated by the utopia point $\bm{v}_u$, and encloses the whole constraint set $\mathcal{G}$. In the second iteration, $\bm{v}_u$ is projected and new vertices are generated as $\bm{v}_1=[0.5 \ 1]$, $\bm{v}_2=[1 \ 0.5]$, $\bm{v}_3=[0.5 \ 0.5]$. In the third iteration, we assume that $\bm{v}_2$ minimizes the objective function, thus new vertices $\bm{v}_4=[0.66 \ 0.5]$, $\bm{v}_5=[1 \ 0.33]$, $\bm{v}_6=[0.66 \ 0.33]$ are generated by its projection. The last subfigure shows the $6$th iteration, where $\mathcal{P}_6$ encloses the constraint set more tightly, and new vertices are concentrated around the optimal solution $\bm{v}^*$.

\begin{figure}[t]
    \begin{center}
    \subfigure[]{\includegraphics[width=0.44\linewidth]{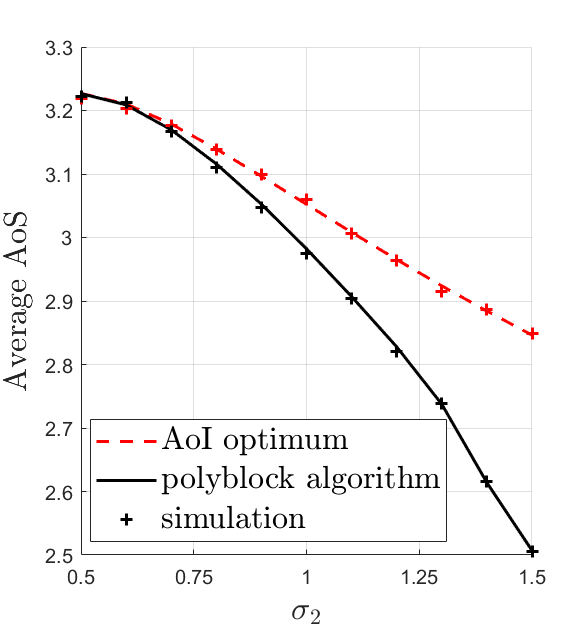}}
    \subfigure[]{\includegraphics[width=0.44\linewidth]{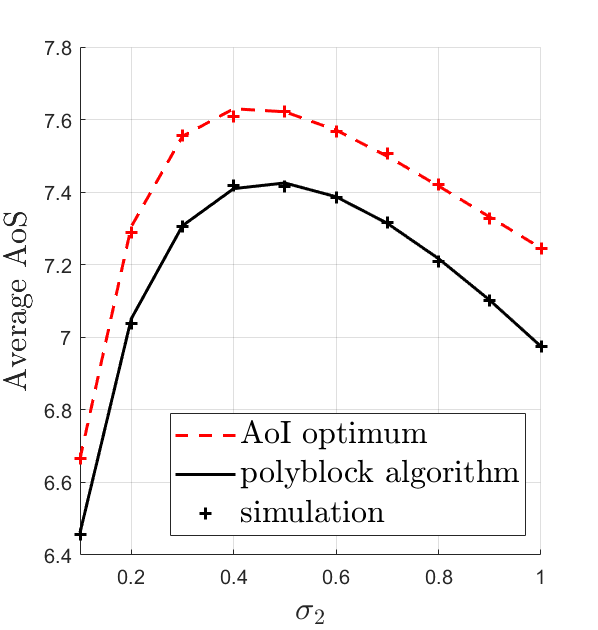}} 
    \end{center}  
    \centering    
    \vspace{-0.3cm}
    \caption{Comparison of average AoS values between the proposed polyblock algorithm and AoI optimum sampling rates for scenarios: (a) $\bm{n}=[8 \ 4]$, $\bm{\sigma}=[1.5 \ \sigma_2]$, and (b) $\bm{n}=[4 \ 6 \ 8]$, $\bm{\sigma}=[0.5 \ \sigma_2 \ 0.5]$.}
    \label{fig:vs}
\end{figure}

\begin{figure}[t]
    \centering
    \includegraphics[width=0.95\linewidth]{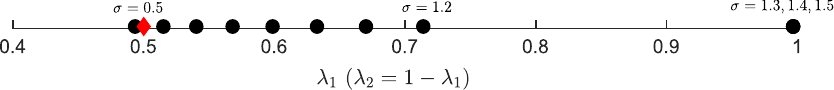}
    \caption{Near-optimal $\lambda_1$ values for varying $\sigma_2$ for scenario a.}
    \label{fig:values}
    \vspace*{-0.3cm}
\end{figure}

\section{Numerical Results}
We verify our analysis and proposed optimization methods with numerical results in Fig.~\ref{fig:vs} for two scenarios:(a) $\bm{n}=[8 \ 4]$, $\bm{\sigma}=[1.5 \ \sigma_2]$, and (b) $\bm{n}=[4 \ 6 \ 8]$, $\bm{\sigma}=[0.5 \ \sigma_2 \ 0.5]$ using varying $\sigma_2$ that vary in $0.1$ and $0.25$ increments for scenarios (a) and (b), respectively. We chose the parameter of the polyblock algorithm as $\epsilon=10^{-3}, \ \delta=10^{-5}$ and $\epsilon=10^{-2},\ \delta=10^{-4}$ for scenarios (a) and (b), respectively. To compare the performance, we use the AoI-optimal policy as a benchmark policy. Since AoI metric ignores the source dynamics, it is minimized if the sampling budget is allocated equally among the sources, i.e, $\lambda_k=\frac{1}{K}$. In these simulations, lines are obtained by our analysis, and simulation results for the same scenarios are marked by the symbol `$+$'. The match between them verifies our analysis. In addition, we observe that allocating sampling rates to minimize AoI results in a significantly higher AoS. 

As the intuition and \eqref{eq:ES} suggest that if a process has a large state space (low $\rho$), or changes slowly (low $\sigma$), a larger sampling rate ($\lambda$) is required to reach the same AoS. In addition, the AoS is lower bounded as $\lambda$ increases, because having a constant and arbitrary estimation for the process periodically resets the AoS from the ergodicity of the process. In Fig.~\ref{fig:values}, we illustrate the optimum $\lambda_1$ values (marked by black circles) alongside the AoI optimum sampling rate (marked by a red diamond). As expected, as $\sigma_2$ increases, the optimum sampling rate $\lambda_1$ gets larger, and eventually it converges to only sample source $1$ for large $\sigma_2$ values. Since the polyblock algorithm is near-optimal, it is terminated before $\lambda_1$ reaches $1$. We additionally obtain a sampling rate that maximizes BF, and observe the \emph{winner takes all} structure, which results far away from the optimal solution in most cases. Particularly, BF is maximized by $\bm{\lambda}=[1\ 0]$ and $\bm{\lambda}=[0\ 1]$ for $\sigma_2\leq1.1$ and $\sigma_2\geq1.2$, respectively.

\newpage
\clearpage

\bibliographystyle{IEEEtran}
\bibliography{bibl}

\end{document}